\documentclass[prl,amssymb,amsmath,superscriptaddress,preprint]{revtex4-1}
\usepackage{graphicx}
\usepackage{dcolumn}
\usepackage{bm}
\usepackage{color}
\usepackage[mathlines]{lineno}
\begin{document}
\title{Local versus global stretched mechanical response in a supercooled liquid near the glass transition}
\author{Baoshuang Shang}
\affiliation{Beijing Computational Science Research Center, Beijing 100193,  China}
\affiliation{Univ. Grenoble Alpes, CNRS, LIPhy, 38000 Grenoble, France}
\author{J\"{o}rg Rottler}%
\affiliation{%
 Department of Physics and Astronomy and Quantum Matter Institute,
University of British Columbia, Vancouver BC V6T 1Z1, Canada
}
\author{Pengfei Guan}
\email{pguan@csrc.ac.cn}
\affiliation{Beijing Computational Science Research Center, Beijing 100193, China}
\author{Jean-Louis Barrat}
\email{jean-louis.barrat@univ-grenoble-alpes.fr}
\affiliation{Univ. Grenoble Alpes, CNRS, LIPhy, 38000 Grenoble, France}
\date{\today}
\begin{abstract}
Amorphous materials have {a rich relaxation spectrum}, which is usually described in terms of a hierarchy of relaxation mechanisms. In this work, we investigate the local dynamic modulus spectra in a model glass just above the glass transition temperature by performing a mechanical spectroscopy analysis with molecular dynamics simulations. We find that the spectra, at the local as well as on the global scale, can be well described by the Cole-Davidson formula in the frequency range explored with simulations. Surprisingly,  the Cole-Davidson stretching exponent does not change with the size of the local region that is probed. The local relaxation time displays a broad distribution,  as expected  based on dynamic heterogeneity concepts, but the stretching is obtained independently of this distribution. We find that the size dependence of the local relaxation time and moduli can be well explained by the elastic shoving model.
\end{abstract}
\maketitle
Nonexponential or stretched exponential relaxation is ubiquitous in amorphous materials, and is recognized as one of the key features in supercooled liquid and glassy states \cite{phillips1996stretched,RevModPhys.78.953}.  It appears in {many relaxation processes} at equilibrium or out of equilibrium, such as aging, stress relaxation and dielectric or mechanical relaxation spectra \cite{gotze1992relaxation,Wang2012a}.
However, the origin of the stretching is still controversial \cite{cavagna2009supercool}. Two hypotheses are typically put forward to explain the stretching:  one identifies the stretched relaxation as resulting from dynamic heterogeneity in different regions of space, the other assumes that the relaxation in amorphous material is uniform, with stretched relaxation being a local feature \cite{ediger1996supercooled,richert2002heterogeneous}.

These different views can to some extent be reconciled within the now widely accepted concept of dynamical heterogeneity, which has been confirmed both in experiment and molecular simulation \cite{berthier2011overview}. The supercooled liquid, for example, can be separated into fast regions of high mobility and slow regions with lower mobility, with a ``slow" or ``fast" character that persists over times comparable to the total $\alpha$ relaxation time. Mathematically, stretched exponential relaxation can be described as a superposition  of simple exponential relaxation processes \cite{edholm2000stretched}. It is then a natural hypothesis to assume that the slow and fast regions associated with dynamical heterogeneity each have a simple exponential relaxation, and that the global stretching results from the different relaxation times associated with different regions, which may be broadly distributed. In fact, this natural assumption was recently formalized in a series of works by Masurel {\it et al.} \cite{masurel2015role,PhysRevLett.118.047801,masurel2017dynamical}, who developed a mesoscale  model to describe the  viscoelastic spectrum in a polymer model near the glass transition temperature. In their model, every local region is described as a single Maxwell Voigt element, with a single relaxation time assigned randomly from a broad (log normal) distribution. {Based on the idea that dynamic and elastic heterogeneity are related, Schirmacher\cite{PhysRevLett.115.015901} also uses a local Maxwell model to describe the relaxation spectra within a mean field theory.}

However,  this assumption that the stretched exponential relaxation arises from  simple exponential relaxation in  local regions has not, to our knowledge, been proven in direct investigations. Only indirect consequences, as in the work of Masurel {\it et al,}  have been explored. In this work, we question directly the validity of this assumption for mechanical properties, using the flexibility offered by molecular dynamics simulations. We build on previous explorations of static properties such as local elastic constants \cite{PhysRevE.87.042306} or thermomechanical \cite{Baoshuang2018} properties and develop a methodology that allows us to obtain the dynamic modulus spectrum in a supercooled liquid near the glass transition at different length scales.  We find that different dynamical spectra can be well fitted by a Cole-Davidson expression, with a distribution of relaxation times that evolves with the measurement scale. However, surprisingly, the stretching exponent does not change with increasing the spatial scale, and is nontrivial at the smallest scale investigated.
Furthermore, We find a strong correlation between the local modulus and relaxation time, which can be rationalized  within an elastic shoving model \cite{RevModPhys.78.953} at the local scale, and the size dependence of the average relaxation time and shear modulus can be well explained by confinement effects, which reflect the nature of elastic interactions in supercooled liquids.

\begin{figure}[t]
    \centering
    \includegraphics[width=0.8\columnwidth]{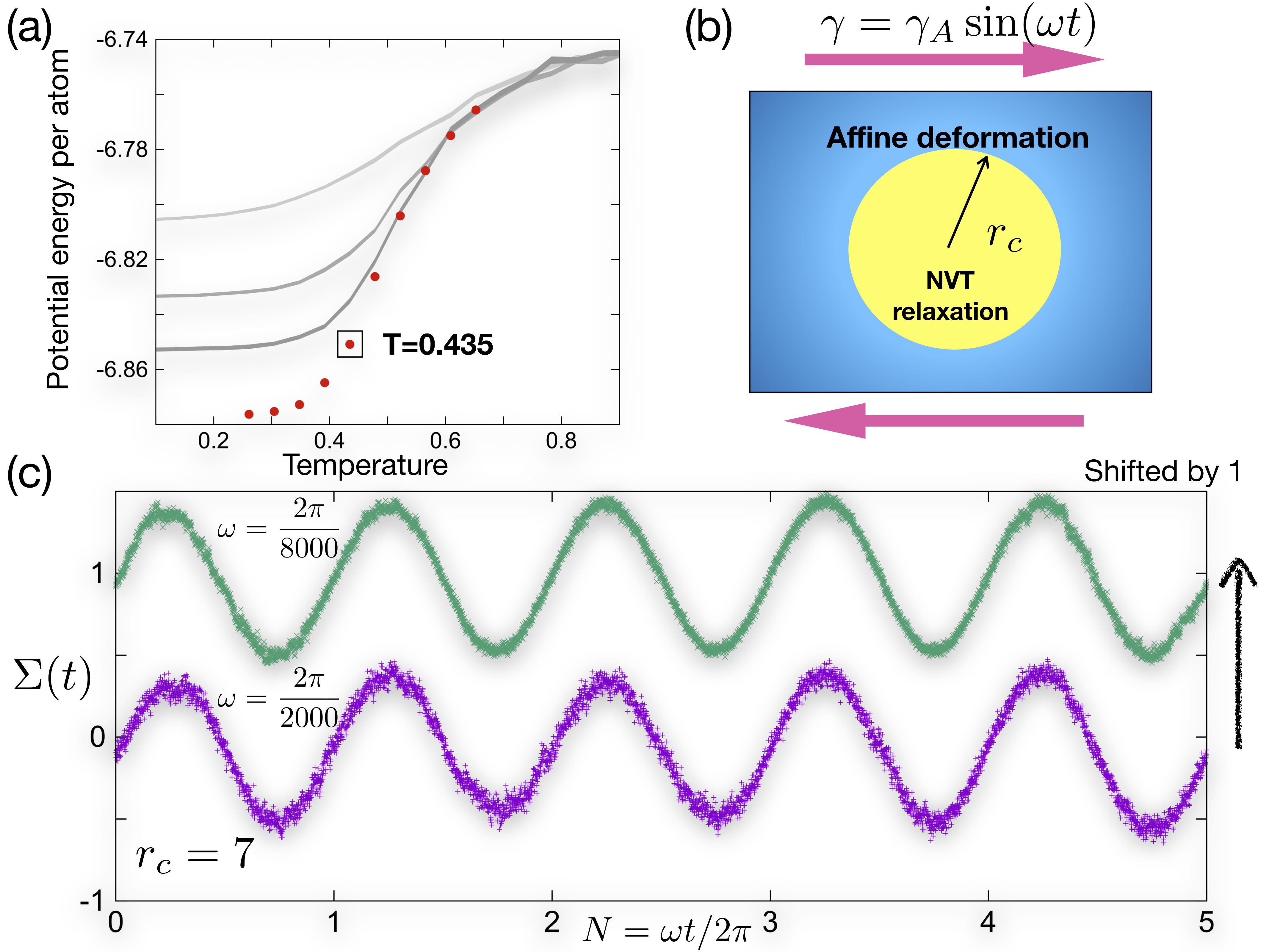}
    \caption{\textbf{Local mechanical spectroscopy analysis.} (a) Inherent structure energy versus temperature during the quench process. The solid lines from top to bottom represent three different quenching rates: $4.35 \times 10^{-3}$, $4.35 \times 10^{-4}$, $4.35 \times 10^{-5}$. Solid points represent the quenching rate $1.67 \times 10^{-6}$. The state used for the  mechanical spectroscopy analysis in this work is shown by the box, at a temperature  T=0.435, close  to the mode coupling temperature of the model \cite{PhysRevLett.73.1376}. (b) Schematic representation  of the local mechanical spectroscopy analysis for a selected region ($r_c$) with sinusoidal strain loading. (c) Local stress response of a region of size $r_c=7$ for two different frequencies shifted by one for clarity (upper:$\omega={2\pi}/{8000}$,  lower: $\omega={2\pi}/{2000}$) in the first 5 cycles.}
    \label{fig:1}
\end{figure}

\begin{figure}[t]
    \centering
    \includegraphics[width=0.8\columnwidth]{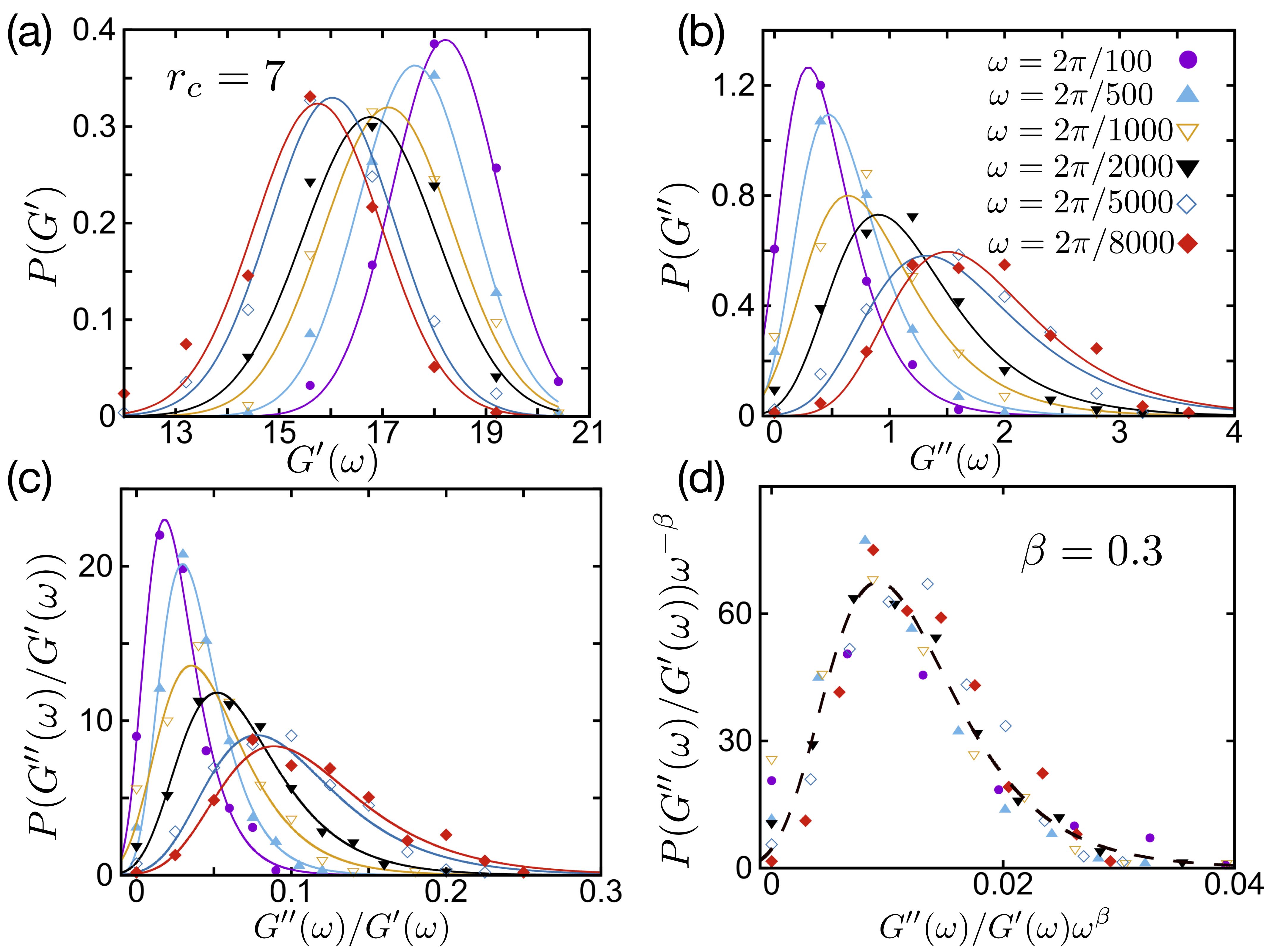}
    \caption{\textbf{Probability distribution of local dynamic moduli for} $r_c= 7$. (a), (b): probability distribution of local storage modulus $G'(\omega)$ and local loss modulus $G''(\omega)$ for different frequencies. (c) probability distribution of loss and storage modulus ratio $\frac{G''(\omega)}{G'(\omega)}$, (d) collapse of the probability distribution in (c) by rescaling the data with $\omega^{\beta}$ for different loading frequency (see text and Fig.~\ref{fig:4} for values of $\beta$). Solid line in (a) is a Gaussian distribution, the solid line in (b), (c) and dashed line in (d) are Gumbel distributions \cite{gumbel1954statistical}. Note that there is no particular  physical reason to choose those distributions except to capture the trend of the data with loading frequency.}
    \label{fig:2}
\end{figure}

\begin{figure}[t]
    \centering
    \includegraphics[width=0.8\columnwidth]{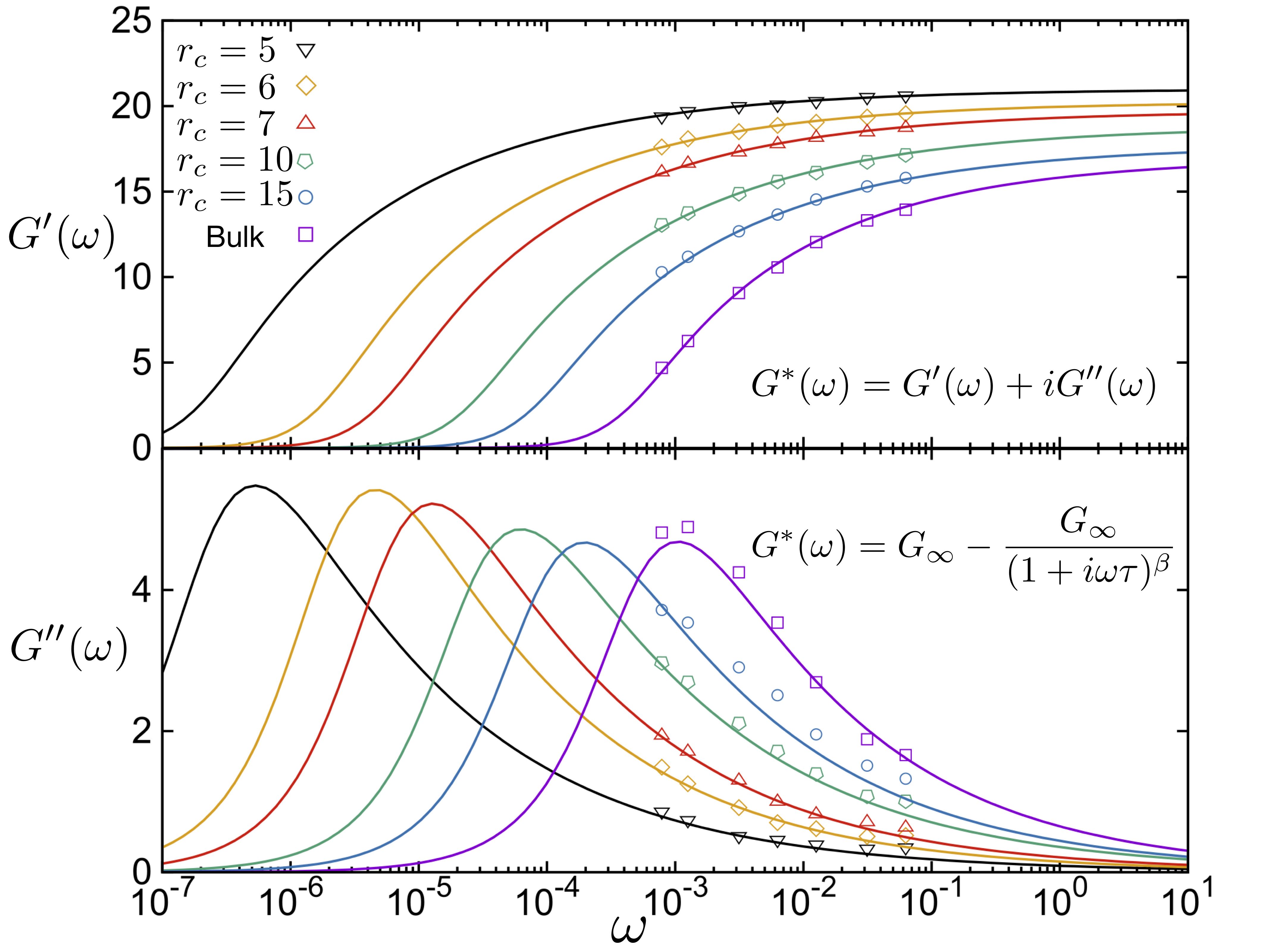}
    \caption{\textbf{Dynamic moduli spectra for different sizes.} The upper panel shows the storage moduli the lower panel the loss moduli, and the solid lines are a fit to the modified Cole-Davidson formula. Error bars are of order the symbol size and are obtained by averaging over different numbers of local regions, for $r_c=5$ : 512 regions, $r_c=6,7,10$ : 216 regions, $r_c=15$ : 64 regions, for bulk sample : 10 independent samples. }
    \label{fig:3}
\end{figure}
Our study of local viscoelastic properties is based on molecular dynamics (MD) simulations of a classical 80:20 binary Lennard-Jones (LJ) glass model \cite{PhysRevE.52.4134} using the LAMMPS package \cite{plimpton1995fast}. The interaction potential was truncated and force shifted \cite{toxvaerd2011communication}. LJ units based on the mass $m$, the interaction diameter  $\sigma$ and  energy $\epsilon$  of the large particles are used throughout, with a  time unit  $\sqrt{m\sigma^2/\epsilon}$. Ten independent simulation samples containing 80,000 atoms each were generated to improve statistics. Fig.~\ref{fig:1}(a) illustrates the potential energy per particle during quenching of the sample at various rates and constant volume in periodic boundary conditions. The quench rate used in our study is $1.67 \times 10^{-6}$. The temperature was controlled  with a Berendsen thermostat \cite{berendsen1984molecular}, the time step was set to 0.001, and the number density $\rho$ was fixed at 1.2. 

To obtain the dynamic shear modulus of the whole sample (bulk) and of local regions for different  frequencies, we performed a numerical analog of a mechanical spectroscopy experiment \cite{PhysRevB.90.144201} at a fix temperature $T=0.435$. To determine the  local properties, we used a modified version of the frozen matrix method, which was previously used to measure local moduli \cite{PhysRevE.87.042306} or local yield stress \cite{PhysRevLett.117.045501} at zero temperature. As shown in Figure \ref{fig:1}(b), we first choose a spherical region of radius $r_c$, then shear the whole sample with sinusoidal strain $\gamma=\gamma_A \sin(\omega t)$ and only let the selected region relax under NVT conditions, while the outside region was affinely deformed according to the imposed sinusoidal shear strain. As a result, the local shear stress $\Sigma (t)$  acquires an oscillatory component at the same frequency, which is extracted from the noise using the numerical analog of a lock-in amplification. As shown in Fig.~\ref{fig:1}(c), the storage shear modulus $G'(\omega)$ and loss shear modulus $G''(\omega)$ were derived from the local stress signal $\Sigma(t)$ using:
\begin{eqnarray}
    I &=\int_{0}^{N \frac{2\pi}{\omega}} \sin(\omega t)\frac{\Sigma(t)}{\gamma_A} dt \\
    J &=\int_{0}^{N \frac{2\pi}{\omega}} \cos(\omega t)\frac{\Sigma(t)}{\gamma_A} dt 
\end{eqnarray}
\begin{equation}
    G'(\omega) =\frac{I\omega}{N \pi}; \ \ 
    G''(\omega) =\frac{J\omega}{N \pi}
\end{equation}
\begin{equation}
   G^{*}(\omega) \equiv G'(\omega)+iG''(\omega)
\end{equation}
where $N$ is the number of strain cycles and $\gamma_A$ is the amplitude of strain. Here we choose $N=5$, $\gamma_A=0.02$; the influence of the choice of $N$ and $\gamma_A$ is discussed in Supplementary Material (SM)\cite{supp}. In order to sample the space, the center points of the free regions were selected from a regular grid. For $r_c=5$ , the sample was meshed as $8 \times 8 \times 8$, for $r_c=6, 7, 10$, the grid was $6 \times 6 \times 6$ and for $r_c=15$, the grid was $4 \times 4 \times 4$. The bulk dynamic modulus was obtained by applying an oscillatory strain to the whole simulation box.

 For a series of different frequencies, a probability distribution of the storage and loss moduli can be obtained from the statistics over different zones.  As shown in Fig.~\ref{fig:2}(a) and (b), this probability distribution of the local moduli shows a  distinct  frequency dependence. As the frequency of the loading increases, the most probable value of the local storage modulus shifts to higher values and the one of the local loss modulus shifts to lower values.  This  trend is general, independent of the size of the  local region  (for different  sizes see Fig. S2 for $r_c=5$, Fig. S3 for $r_c=6$, Fig. S4 for $r_c=10$ in the SM).

In Fig. \ref{fig:3},  the average values of the local storage and loss modulus (obtained from the probability distribution function) are now plotted as a function of frequency and compared with the bulk values obtained from dynamical mechanical analysis on the whole sample.   Both storage and loss moduli are notably influenced by size. However, all  dynamic moduli frequency spectra, regardless of size, can be well fitted by a Cole-Davidson form \cite{davidson1951dielectric}, 
\begin{equation}
G^{*}(\omega) =G_{\infty}-{G_{\infty}}{(1+i\omega \tau)^{-\beta}},
\label{cd-eq}
\end{equation}
where $i=\sqrt{-1}$, $G_{\infty}$ is the high frequency shear modulus, $\tau$ is the relaxation time, and $\beta$ is Cole-Davidson stretching exponent. Since our simulations are performed in the supercooled liquid state, where we expect $G'(\omega)|_{\omega \to 0}=0$,  the Cole-Davidson formula contains only three independent parameters. The formula reduces to the usual Debye model with a single relaxation time for $\beta=1$. We will see below that this value of $\beta$ is clearly outside the uncertainty range on the fit parameters.

\begin{figure}[t]
    \centering
    \includegraphics[width=0.8\columnwidth]{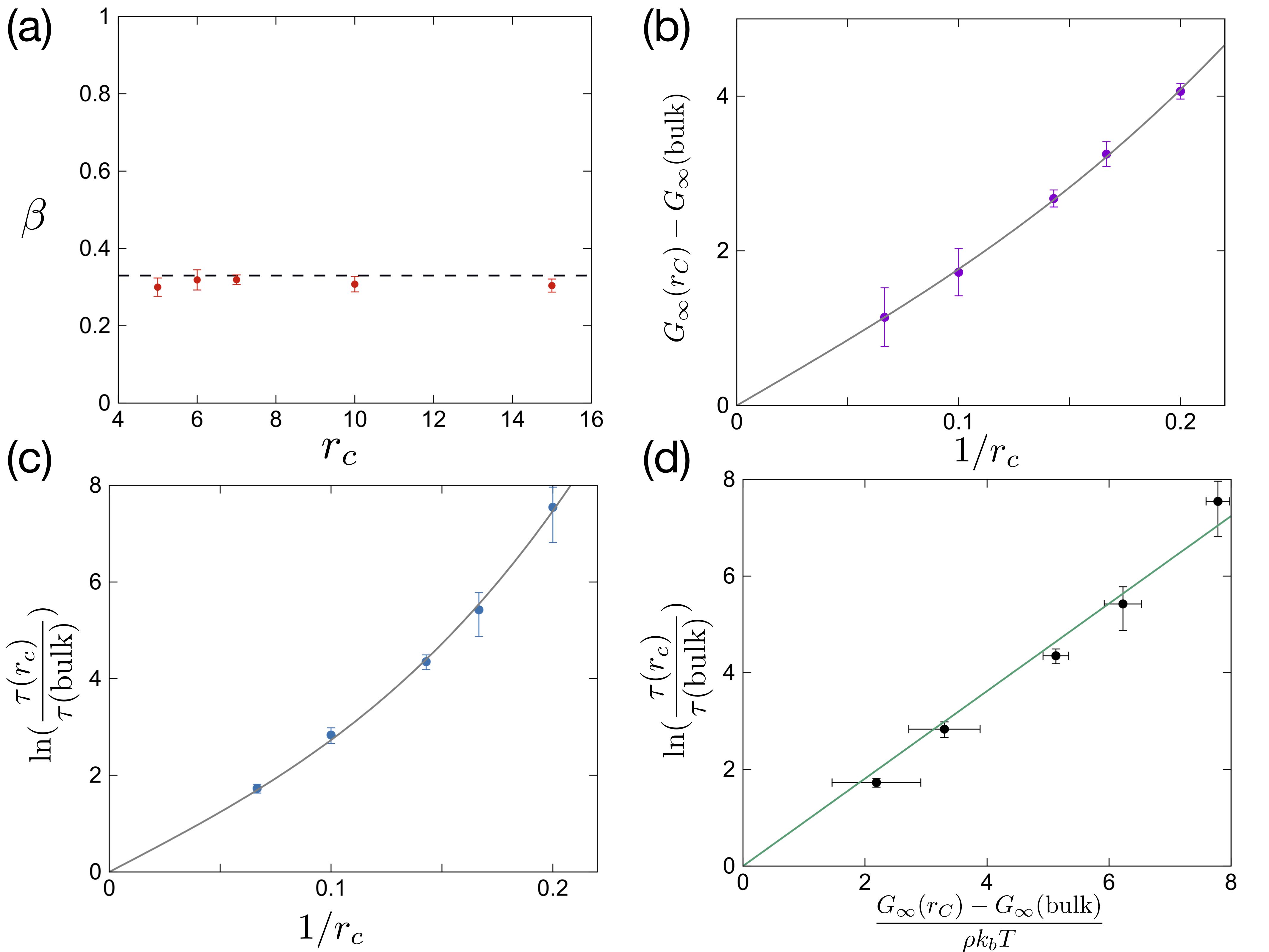}
    \caption{ \textbf{Scale dependence of the heterogeneous relaxation parameters.} (a) Cole-Davidson stretch exponent $\beta$ for different local radii (the dashed line represents the bulk value). (b), (c) local $G_{\infty}$ and local relaxation time $\tau(r_c)$ vs.~inverse confinement radius $1/r_c$. The solid curve is a best fit of the data with the formula $\frac{A}{r_c}+\frac{B}{r_c^3}$ (see text). (d) correlation between local relaxation time $\tau$ and local $G_{\infty}$. Errorbars are obtained from fitting the data in Fig.~\ref{fig:3} within a 95\% confidence interval.} 
    \label{fig:4}
\end{figure}

Figure \ref{fig:4} reports the parameters obtained from fitting the frequency dependent mechanical response, averaged over all sampled regions as shown in Fig.~\ref{fig:3},  as a function of the size of the region. Surprisingly, the stretching exponent $\beta$  does not change with size  (see Fig.~\ref{fig:4}(a)).
However, since the response shown in Fig. \ref{fig:3} is averaged over many different regions, this feature can still be explained  by two different hypotheses: either the average value is the superposition of individual relaxations, and the each individual region still follows a Debye relaxation with $\beta=1$ in the Cole-Davidson formula, or the stretching is a feature of every individual region.

In order to distinguish between these two possibilities, it would in principle be adequate to perform an individual fit of eq.~\eqref{cd-eq} to the frequency dependence of the mechanical response for each region. Unfortunately, such a procedure turns out to be difficult in view of the relatively large statistical uncertainty of individual spectra. Instead, we choose to investigate the consistency of the above hypotheses with the observed {\it statistical} properties of the locally measured $G'(\omega)$ and $G''(\omega)$, as characterized by the probability distribution functions shown in Fig.~\ref{fig:2}. 

To this end, we calculate for every zone and loading frequency the ratio of loss and storage modulus $G''(\omega)/G'(\omega)$. Within a Cole-Davidson model, this ratio reads
\begin{equation}
\frac{G''(\omega)}{G'(\omega)} = \frac{\sin(\beta \theta)}{[1+\omega^2\tau^2]^{\frac{\beta}{2}}-\cos(\beta \theta)},
\end{equation}
where $\theta \equiv \arctan (\omega\tau)$. Considering that the loading frequency is generally such that $\omega \tau \gg 1$,  this formula  can be simplified as
\begin{equation}
\frac{G''(\omega)}{G'(\omega)} \approx \sin(\frac{\pi}{2}\beta)\omega^{-\beta}\tau^{-\beta}.
\label{eq:ratio1}
\end{equation}

Figure \ref{fig:2}(c) shows the probability distribution of this ratio for different loading frequencies. As expected from eq.~\eqref{eq:ratio1}, the probability distribution function  is  sensitive to loading frequency, with a most probable value that decreases with increasing frequency.  The width  of the distribution also decreases with  increasing  $\omega$. However, if one now rescales the data to obtain the probability distribution of $G''(\omega)/G'(\omega) \omega^{\beta}$, which according to the Cole Davidson model is $\tau^{-\beta}$,  a very good collapse of the different distributions, independent of frequency,  is obtained, as shown in  Fig. \ref{fig:2}(d). Here the value of $\beta=0.30$ was set to the one obtained from  fitting the average response (see Fig.~\ref{fig:4})(a)), which shows that the average response is  relevant to  describe the statistical properties of individual zones. This  proves that every individual region actually follows the same stretched relaxation process as the bulk sample. Note that the data in Fig.~\ref{fig:2} corresponds to a particular size of the local region, namely $r_c=7$. However, different sizes lead to similar conclusions\cite{supp}.
 
Following this analysis, one must conclude that the collapsed distributions shown in  Fig.~\ref{fig:2}(d) represent the probability distribution of  $\tau^{-\beta}$  (up to a factor  $\sin(\frac{\pi}{2}\beta)$). From the present data, it follows that the  distribution of $\tau^{-\beta}$ is relatively narrow and {it would slightly increase with $r_c$ within our investigation regime (shown in Fig.~S5 in SM)}, in contrast with the common assumption of a broadly distributed relaxation time, {and the width of the local relaxation time would decrease with $r_c$\cite{supp}.}
As expected from the dynamic heterogeneity picture, relaxation is heterogeneous, but the heterogeneity does not explain the stretching of the relaxation, which is present at the local scale,  and rather homogeneous.

Another interesting conclusion from the analysis shown in figures \ref{fig:3} and \ref{fig:4} concerns the size dependence of the average relaxation times and high frequency moduli. In contrast to the result for the stretching exponent, the size effect is very pronounced for these quantities. We 
explain this dependence by assuming that the relaxation time $\tau$ is related to a local free energy barrier  $\Delta F$ through an Arrhenius law, $\tau \sim e^{\frac{\Delta F}{k_b T}}$.
Assuming further, following the general ideas of elastic ``shoving" models \cite{RevModPhys.78.953}, that the free energy barrier is mainly due to  elastic energy, then the high frequency modulus $G_{\infty}$ is directly correlated with  the free energy barrier, as $G_{\infty} \sim \Delta F$. The consistency of these  assumptions can be directly checked using a parametric plot of  $\ln \tau$ versus $G_{\infty}$ with the size of the region as a parameter. Such a plot is shown in figure \ref{fig:4}(d) and reveals an excellent correlation between these quantities.

In order to understand the origin of the size effect on the energy barrier, we may assume that this energy barrier is associated with shear transformations taking place within the zone, and compare the situation in which the outside region is affinely deformed with  the one in which this region is allowed to relax. Two effects will contribute to increasing the elastic energy in the frozen matrix configuration. Firstly, the local shear modulus in the vicinity of the frozen boundary will have a smaller nonaffine contribution, as nonaffine relaxation is partially prevented by the boundary. It is well known that the nonaffine contributions decrease the shear modulus compared to the purely affine (Born) value \cite{PhysRevLett.93.175501},
so that a shell in the vicinity of the boundary will effectively have an increased modulus compared to the bulk. Such a surface contribution is expected to contribute as $1/r_c$ to the shear modulus of the zone. In addition, since the shear transformations in supercooled liquid shows a Eshelby inclusion-like pattern \cite{illing2016strain,PhysRevLett.113.245702},the energy barrier of shear transformations can be represented by the Eshelby external field energy\cite{eshelby1957determination,picard2004elastic}. For the local region,   the displacement field associated with a shear transformation has to vanish at the boundary $r_c$. This effectively increases the average shear strain, and an order of magnitude estimate leads to an increase in the elastic energy scaling as $1/r_c^3$ due to this effect \cite{li2007eshelby}. As a result, we propose to fit the size dependence of the shear modulus (or equivalently of the energy barrier) using the form:
\begin{equation}
 G_\infty(r_c)  = G_\infty(\text{bulk}) (1+ \frac{A}{r_c}+\frac{B}{r^3_c})
 \label{eqn:form}
\end{equation}
The result of this fit yields  $A \simeq 1.0 $, $B\simeq 5.4$. In the absence of any other relevant length scale, these values are consistent with the particle size and volume, respectively. 
Together with the correlation between relaxation time and shear modulus shown in Fig.~\ref{fig:4}(d), this fits provide evidence for the elastic nature of the local energy barriers. 

 We have investigated the scale and frequency dependence of the viscoelastic moduli  of a generic  glass forming system near the glass transition. In the system we investigated, the stretching of the relaxation  cannot be simply assigned to  the superposition of dynamic heterogeneities, but already exists at a very local scale. The smallest size we investigated,   $r_c = 5$, contains a few hundreds atoms. It corresponds to the typical size below which the mechanical response of a local region no longer follows Hooke's law \cite{PhysRevE.80.026112}, and can be considered as the possible starting point of a coarse graining approach.
 However, the complexity of this ``elementary volume" is already such, that a naive coarse graining starting from a simple Maxwell description at the local scale is not possible. In fact, this result is consistent with established results concerning the  potential energy landscape of systems with a small number of particles \cite{doliwa2003finite,rehwald2012coupled}, which already for a few tenths of particles has a complexity comparable to that of larger systems.
 {It can also be understood by considering the actual length scales involved in the problem. Recent theories of glasses introduce the dynamical length $\xi_{d}$ and the static length $\xi_{s}$. $\xi_d$ quantifies the fact that, in a bulk system, two regions at points $r$ and $r^\prime$ have different mobilities at times $t$ and $t^\prime$. This length grows rather rapidly near the glass transition, and is comparable to $r_c=5$ for our system, according to earlier work \cite{karmakar2014growing}. In contrast,  $\xi_s$ characterizes the amorphous order, grows much more modestly in the temperature range we are studying \cite{karmakar2014growing}, and is less than 3 particle diameters. As a result, the smallest value of $r_c\approx 5$ at which we can define a shear modulus already contains several amorphous regions in the spirit of random first order theory \cite{Wolynesbook}. In view of this ordering of the relevant length scales, the elastic barrier defined on the scale $r_c$ is not able to probe directly the expected growth of $\xi_d$. One may however speculate, as pointed out by Bouchaud and Biroli in chapter 2 of \cite{Wolynesbook}, that the relevant volume in the elastic barrier involves $\xi_d^3$ rather than the atomic volume.}
 
 While the complexity that determines the stretching exponent is essentially insensitive to the scale, a nontrivial scale dependence emerges due to the elastic nature of energy barriers  that govern relaxation in supercooled liquids \cite{PhysRevLett.113.245702,PhysRevLett.78.2020}.
 {As discussed above, this dominance of the  elastic aspect appears as we are operating on a scale larger than the  static correlation length \cite{PhysRevLett.119.195501}. A different regime may be observed for $r_c/\xi_s<1$, a regime that we are not able to probe.}
 Understanding the global relaxation on the basis of a coarse graining approach between elastically interacting elements seems therefore a promising approach. It would, however, require to model each element by a complex behavior, an approach which, to our knowledge has not been attempted until now, and may be challenging within the framework of classical finite element codes.

This work is supported by (BSS and PGF) the NSF of China
(Grant Nos.51601009,Nos.51571011), the MOST 973 Program (No.2015CB856800) and the NSAF joint program (No.U1530401). BSS and PFG acknowledge the computational support from the Beijing Computational Science Research Center (CSRC). J-L. Barrat is supported by Institut Universitaire de France.

\newpage
\section{Supplementary Material}
\renewcommand{\thefigure}{S\arabic{figure}}
\renewcommand{\theequation}{S\arabic{equation}}
\renewcommand{\thesection}{S\arabic{section}}
\setcounter{figure}{0}
\setcounter{equation}{0}
\setcounter{section}{0}
\section{Robustness of the calculation of local dynamic modulus}
\begin{figure*}
    \centering
    \includegraphics[width=0.8\columnwidth]{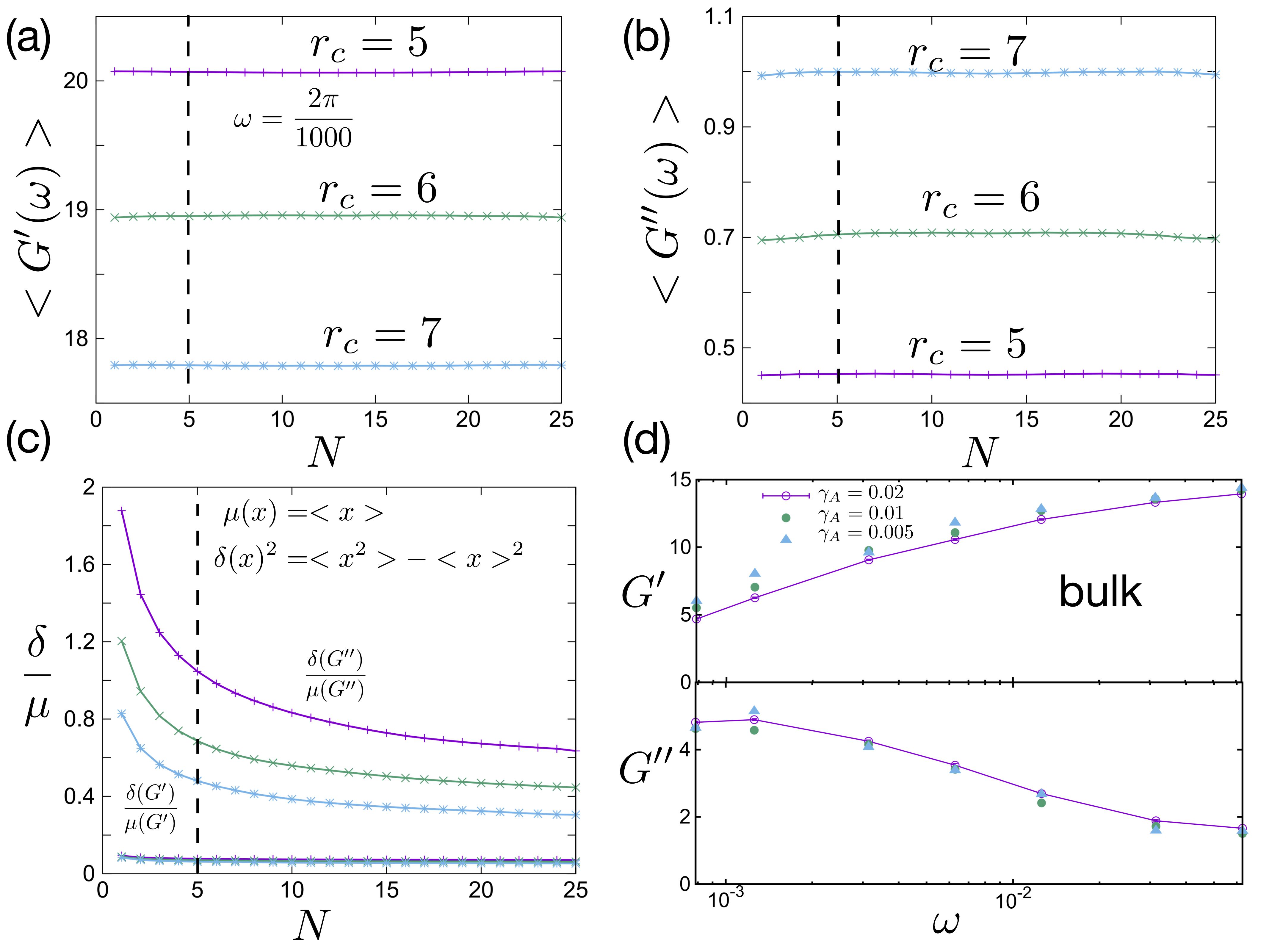}
    \caption{\textbf{dynamical modulus for different cycle numbers and strain amplitudes.} (a),(b) Change in the average value of local storage and loss modulus changes with  number of cycles. (c) Relative standard deviation versus  number of cycles  for the storage and loss modulus. (d) Bulk dynamic modulus frequency spectrum for different strain amplitude, the storage and loss modulus are shown in upper and lower panel, respectively.}
    \label{fig:A1}
\end{figure*}

We have investigated the influence of the number and amplitudes of strain  cycles. In particular, we checked that for  the selected   amplitude of the strain the response is still linear (see Fig. \ref{fig:A1}). We  also checked that the mean value of the local dynamic modulus was not sensitive to the  number of cycles.  The variance in the probability distribution decreases with the number of cycles, but beyond 5 cycles becomes dominated by the intrinsic disorder rather than by the noise on the measurement. 
\newpage

\section{local dynamic modulus for regions  of size \textbf{$r_c=5, 7, \mathrm{and}\, 10$}}
\begin{figure*}
    \centering
    \includegraphics[width=0.8\columnwidth]{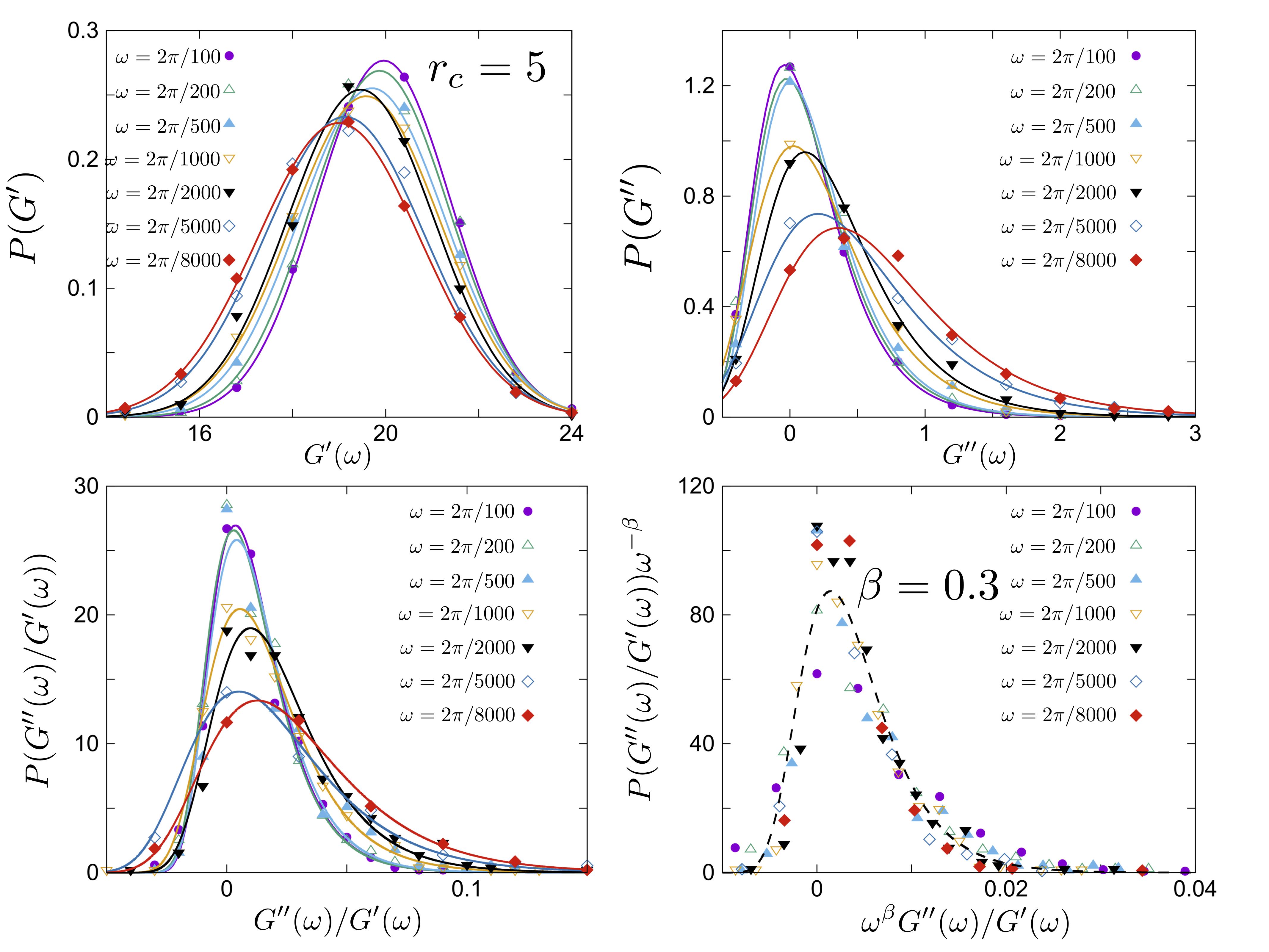}
    \caption{Probability distribution of the local dynamical modulus for $r_c= 5$. (a),(b) probability distribution of the local storage modulus $G'(\omega)$ and local loss modulus $G''(\omega)$ for various  loading frequencies, respectively. (c) probability distribution of loss and storage modulus ratio $\frac{G''(\omega)}{G'(\omega)}$ (d) data collapse of probability distribution in (c) by rescaling the data with $\omega^{\beta}$ for different loading frequencies, $\beta$ is the parameter obtained from Figure 4(a) in the main article. The solid line in (a) is a Gaussian distribution, the solid lines in (b),(c) and dashed line in (d) represent  Gumbel distributions.}
    \label{fig:A2}
\end{figure*}
\begin{figure*}
    \centering
    \includegraphics[width=0.8\columnwidth]{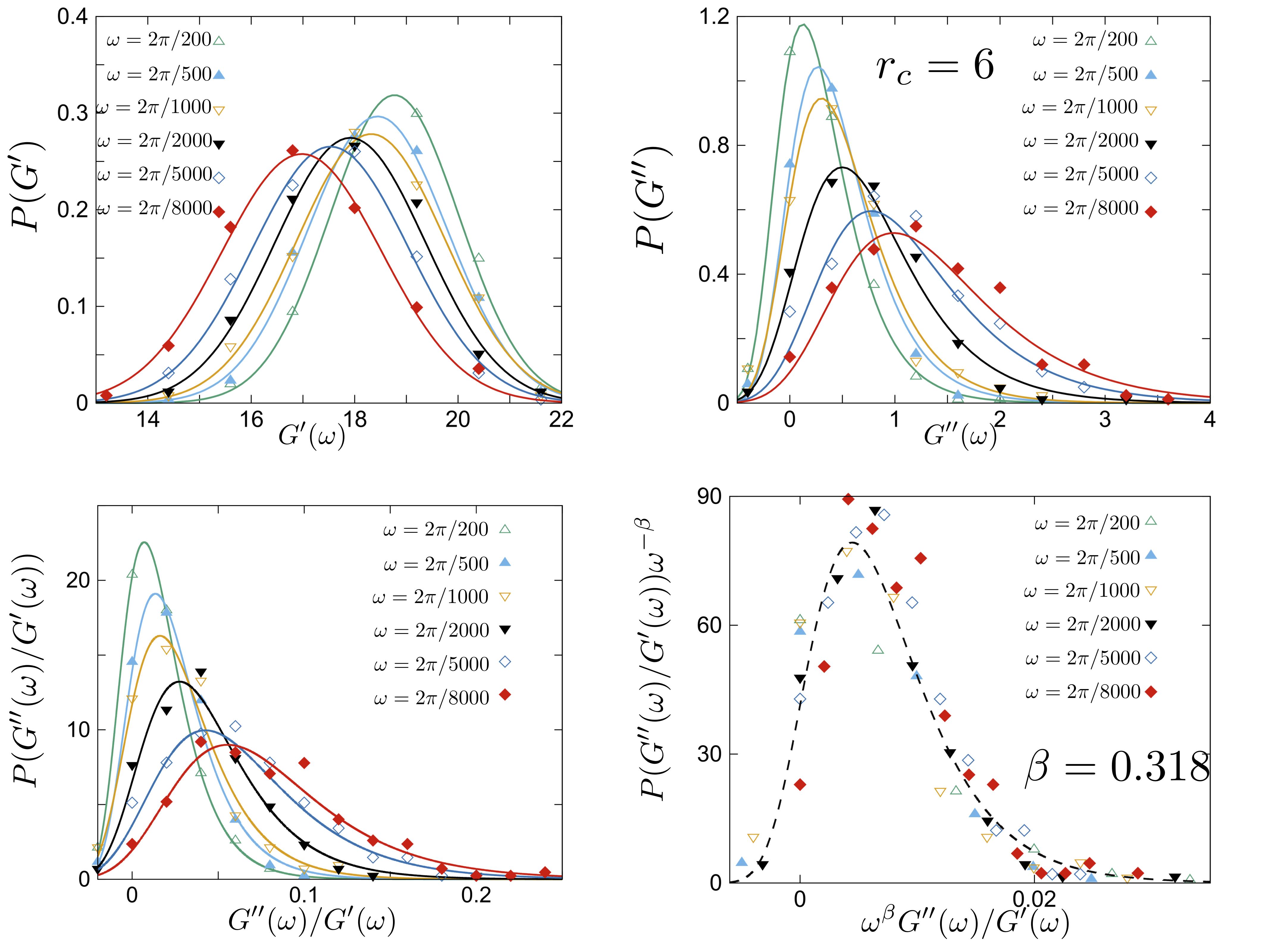}
    \caption{Probability distribution of the local dynamical modulus for $r_c= 6$. (a),(b) probability distribution of the local storage modulus $G'(\omega)$ and local loss modulus $G''(\omega)$ for different loading frequencies. (c) probability distribution of loss and storage modulus ratio $\frac{G''(\omega)}{G'(\omega)}$ (d) data collapse of probability distribution in (c) by rescaling the data with $\omega^{\beta}$ for different loading frequencies, $\beta$ is the parameter from Figure 4(a) in the main article. The solid line in (a) is a Gaussian distribution, solid lines in (b),(c) and dashed line in (d) are Gumbel distributions.}
    \label{fig:A3}
\end{figure*}

\begin{figure*}
    \centering
    \includegraphics[width=0.8\columnwidth]{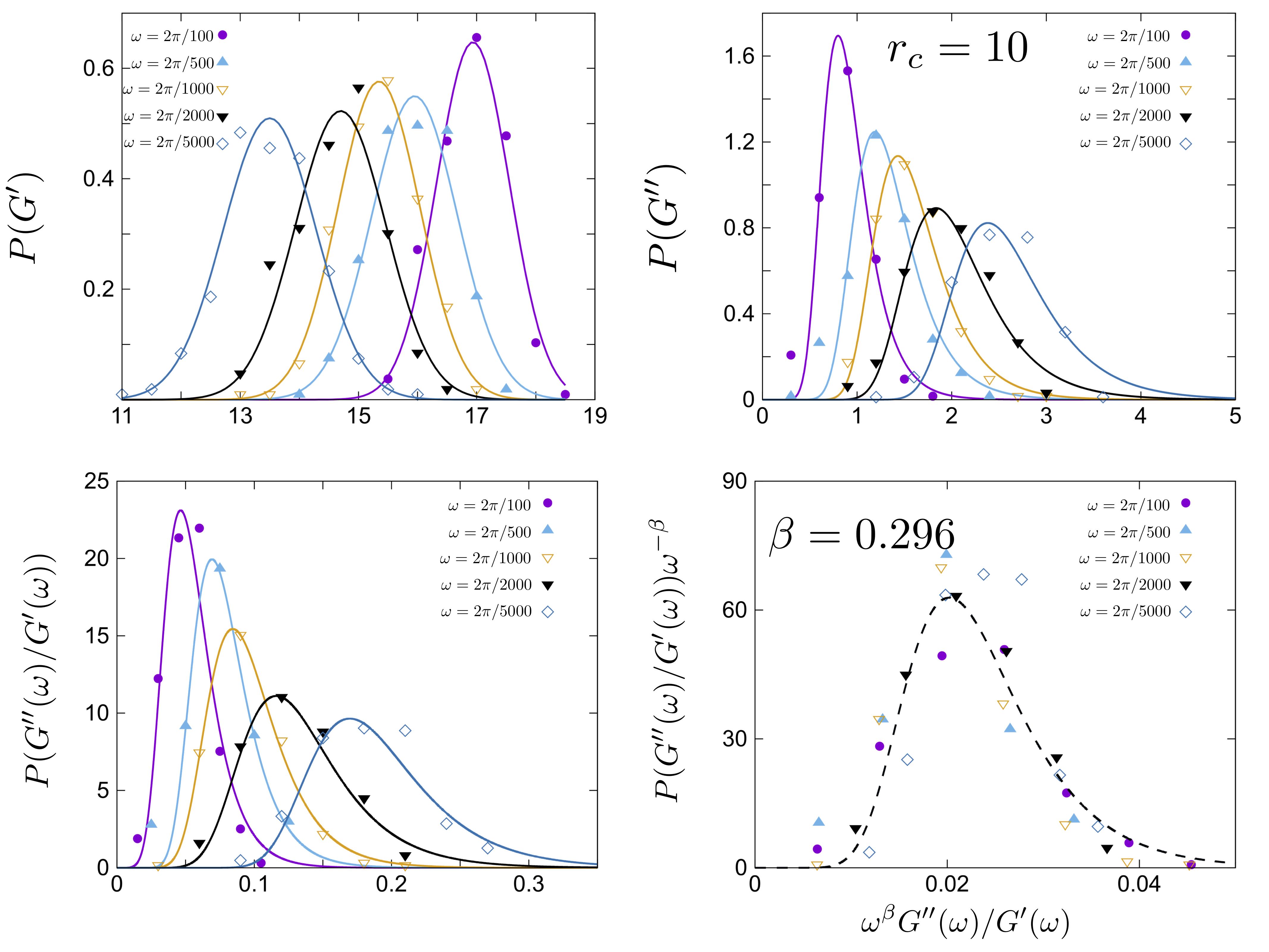}
    \caption{Probability distribution of the local dynamical modulus for $r_c= 10$. (a),(b) probability distribution of the local storage modulus $G'(\omega)$ and local loss modulus $G''(\omega)$ for various loading frequencies. (c) probability distribution of loss and storage modulus ratio $\frac{G''(\omega)}{G'(\omega)}$ (d) data collapse of probability distribution in (c) by rescaling the data with $\omega^{\beta}$ for different loading frequency, $\beta$ is the parameter from Figure 4(a) in the  main article. The solid line in (a) is a Gaussian distribution, solid lines in (b),(c) and dashed lines in (d) are Gumbel distributions.}
    \label{fig:A4}
\end{figure*}

\newpage

\section{Local relaxation time distributions}
First, let us recall that we find that the relaxation time is related to the local shear modulus through:
\begin{equation}
   \tau(r_c)=e^\frac{G_{\infty}(r_c)}{k_{B}T\rho}  
\end{equation} 
If we assume (consistent with many previous findings) that the modulus follows a Gaussian, it follows that the distribution of  $\tau(r_c)$ is a log-normal distribution with the properties:
\begin{equation}
     \langle\ln (\tau(r_c))\rangle = \frac{\langle G_{\infty}(r_c)\rangle}{k_{B}T\rho} \equiv \mu
\end{equation}

\begin{equation}
     \langle\ln (\tau(r_c))^2\rangle- \langle\ln (\tau(r_c))\rangle^2=\frac{SD^2 (G_{\infty}(r_c))}{k_{B}T\rho} \equiv \sigma^2 
\end{equation}
where $SD(x)  \equiv \sqrt {\langle x^2\rangle-\langle x\rangle^2}$ indicates the standard deviation. 
From the properties of the log normal distribution it follows that 
\begin{equation}
 \ln\langle\tau(r_c)\rangle=\mu+\sigma^2/2
 \end{equation}
 \begin{equation}
 SD(\tau(r_c))=\sqrt{e^{\sigma^2}-1}\langle\tau(r_c)\rangle
 \end{equation}
 
\begin{figure*}
    \centering
    \includegraphics[height=0.5\textwidth]{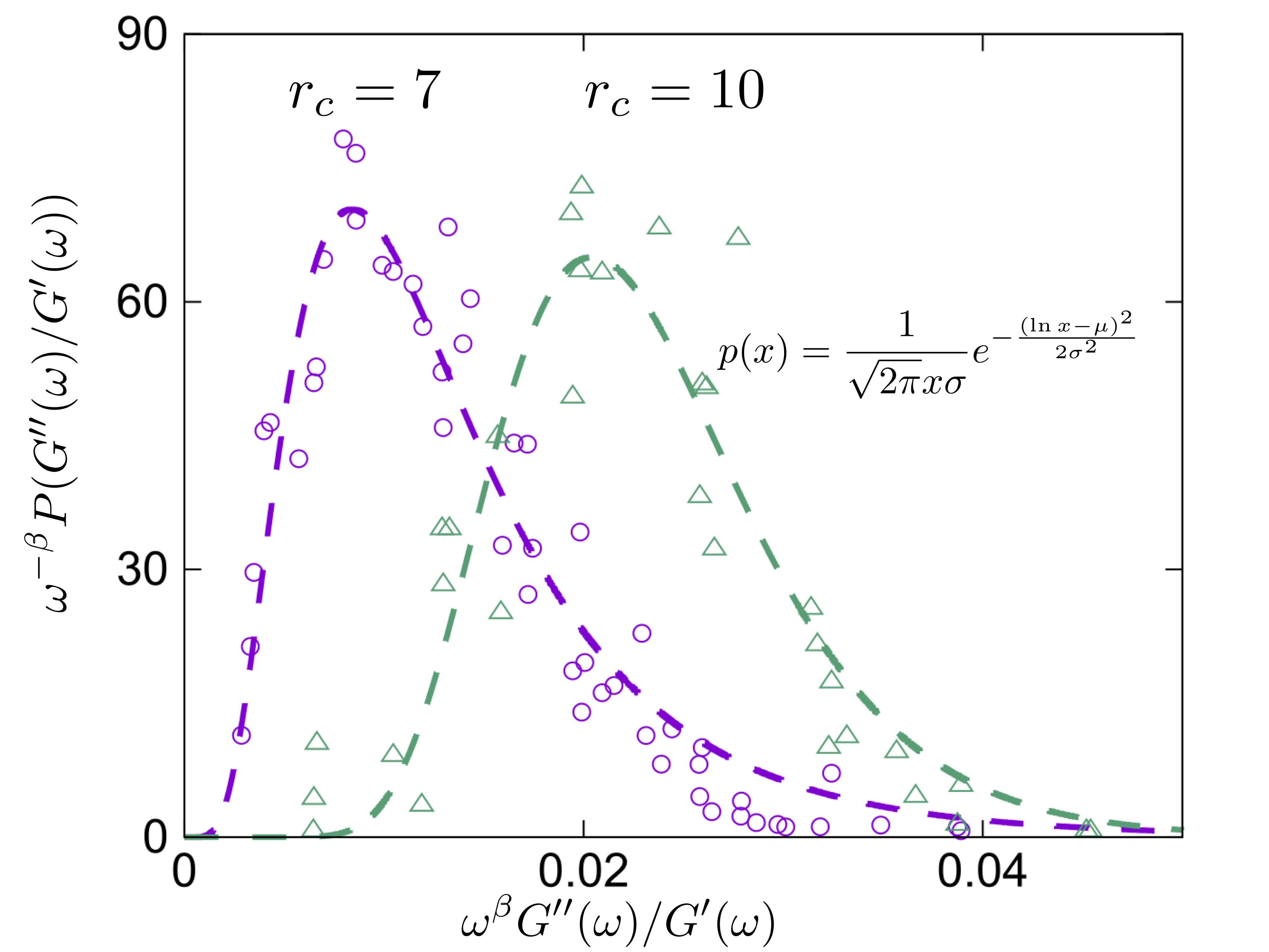}
    \caption{Probability distribution of $\omega^{\beta} G''(\omega)/G'(\omega)$. The data for $r_c=7$ is from Figure 2(d) in main article, $r_c=10$ is from Figure S4(d) in Supplementary material, and the  dashed line is a log-normal distribution for each $r_c$.}
    \label{fig:A5}
\end{figure*}

\begin{figure*}
    \centering
    \includegraphics[height=0.6\textwidth]{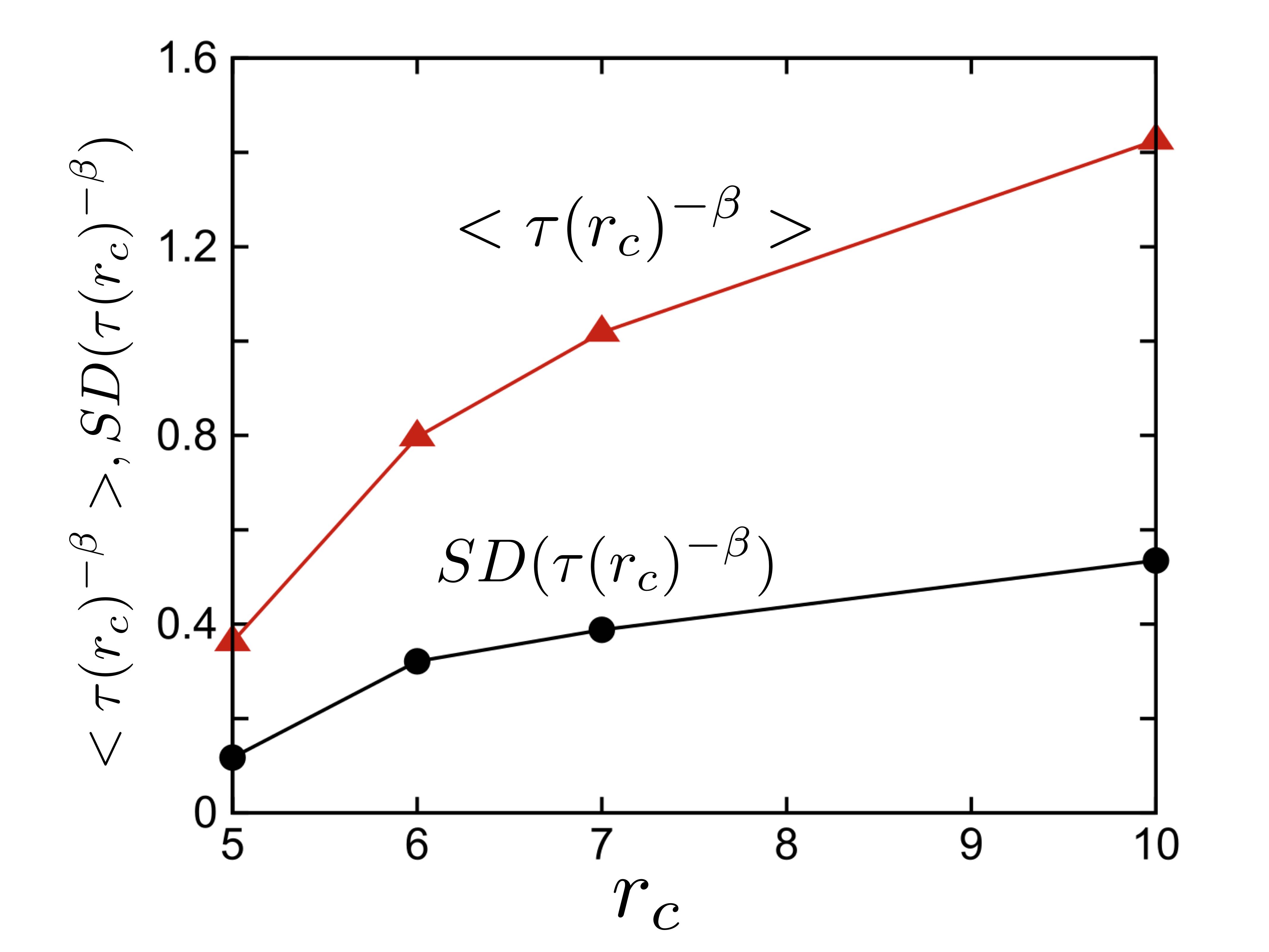}
    \caption{ Mean  and standard deviation of $\tau^{-\beta}$ versus  $r_c$ 
    }
    \label{fig:A6}
\end{figure*}

 The analysis can be extended to the quantity whose distribution is shown in fig 2(d), which is actually $\tau(r_c)^{-\beta}$. This quantity should also follow  a log-normal distribution, shown in Fig.~\ref{fig:A5}, with 
 \begin{equation} 
 \langle\ln\tau(r_c)^{-\beta}\rangle=-\beta \mu
\end{equation}
\begin{equation}
 SD(\ln\tau(r_c)^{-\beta})=\beta \sigma
\end{equation} 
 Again from the properties of the distribution, 
\begin{equation}
 \ln\langle\tau(r_c)^{-\beta}\rangle=-\beta \mu + \beta^2 \sigma^2/2
 \end{equation}
\begin{equation}
SD(\tau(r_c)^{-\beta})=\sqrt{e^{\beta^2\sigma^2}-1}\langle\tau(r_c)^{-\beta}\rangle
\end{equation}
As a result, the standard deviation depends on $r_c$ through the average value $\langle\tau(r_c)^{-\beta}\rangle$, which increases when $r_c$ increases, and the prefactor $\sqrt{e^{\beta^2\sigma^2}-1}$, which is expected to decrease when $r_c$ increases. In the range of values of $r_c$ we have investigated, the net effect is an increase of the standard deviation with $r_c$, as illustrated in Fig.~\ref{fig:A6}. At larger $r_c$, we expect the variation of  $\langle\tau(r_c)^{-\beta}\rangle$ to saturate and the standard deviation to decrease again with $r_c$.
\end{document}